\documentclass[hyper]{JHEP} 
\usepackage{epsfig}
\usepackage{amsmath}
\newcommand\fverb{\setbox\pippobox=\hbox\bgroup\verb}
\newcommand\fverbdo{\egroup\medskip\noindent%
            \fbox{\unhbox\pippobox}\ }

\newcommand\fverbit{\egroup\item[\fbox{\unhbox\pippobox}]}
\newbox\pippobox

\title{Giant Magnon on Deformed $AdS_3\times S^3$}

\preprint{\hepth{}}

\author{
Malak Khouchen and Josef Kluso\v{n} \\
Institute for Theoretical Physics  and Astrophysics\\
Faculty of Science, Masaryk University\\
Kotl\'{a}\v{r}sk\'{a} 2, 611 37, Brno\\
Czech Republic\\
E-mail: \email{malak.khouchen@gmail.com,klu@physics.muni.cz}}

\abstract{We study  giant magnon solutions
 for strings moving on a deformed $AdS_3\times S^3$ background.
  We impose a conformal gauge on the Polyakov action and
  proceed with solving the Virasoro constraints.
   The expressions of the conserved charge $\it{J}$
and the energy of a single magnon excitation are then computed. Then
we determine the dispersion relation of a giant magnon
in the infinite $\it{J}$ limit configuration and we find that for
$\kappa=0$  it reduces to celebrated Hofman-Maldacena dispersion
relation.   }
  \keywords{AdS/CFT correspondence, \  Bosonic
Strings}

\def\th{\tilde{h}}
\def\tf{\tilde{f}}
\def\hT{\hat{T}}

\def\mC{\mathcal{C}}

\begin{document}
\section{Introduction}

String theory in suitable space-time backgrounds can have a holographic description
 in terms of gauge field theories. Type IIB String theory in
  $AdS_5\times S^5$ background has been conjectured to be dual to N=4 super Yang-Mills theory in
4-dimensions \cite{Maldacena:1997re}. The conjectured duality has
passed through various non-trivial tests in the past by analyzing the
spectrum of quantum string states on $AdS_5 \times S^5$ background
and the spectrum of the anomalous dimensions of the N=4 gauge theory
operators in the planar limit. Especially in the semi-classical
approximation, the theory becomes {\it integrable} on both sides
of the duality\footnote{ For reviews of this problems from
different point of views, see
\cite{Minahan:2006sk,Beisert:2004ry,Plefka:2005bk,
Tseytlin:2004xa,Zarembo:2004hp,Swanson:2005wz,Tseytlin:2003ii}.}.
In the plane wave background, string theory is integrable
and its quantization is simple. All string states are generated by
creation oscillators that can be applied to the vacuum
 state to construct all the modes.
 For each of these string states,
  there exist particular dual trace
   operators in the gauge theory \cite{Minahan:2006sk,Berenstein:2002jq}.
    This is one of the AdS/CFT surprising aspects.

It is also well known that we can construct new integrable models
from the given integrable model by applying T-duality
transformations
\cite{Lunin:2005jy,Frolov:2005ty,Ricci:2007eq,Beisert:2008iq}. On
the other hand, a new class of one parameter integrable deformation
of $AdS_5\times S^5$ supercoset model was proposed in
\cite{Delduc:2013qra}, following
\cite{Delduc:2013fga,Klimcik:2002zj,Klimcik:2008eq}. This analysis
was then extended in \cite{Arutyunov:2013ega} where the coordinate
form of the bosonic part of the string action was determined and the
corresponding string metric and the NS-NS two form were found. The
perturbative world-sheet scattering matrix of bosonic particles of
the model was also determined there and it was shown that it agrees
with the  large string tension limit of the q-deformed S-matrix.
Finally, the target space interpretation of a given deformation was
found recently in \cite{Hoare:2014pna} \footnote{For recent
interesting work, see \cite{Kameyama:2014bua}.}. Further, the
deformation of $AdS_3\times S^3$ and $AdS_2\times S^2$ were
introduced here  and it was shown that the corresponding metrics are
direct sums of deformed $AdS_n$ and $S^n$ metrics and are given by
truncations of the corresponding parts of the deformed ten
dimensional metric. Clearly, the integrability of the given models
follows from the integrability of the ten dimensional model.

Many attempts have been undertaken to find out
 the exact spectrum of string theory in Anti-de-Sitter background.
  Though our knowledge of the string spectrum in
  curved backgrounds is still limited,
   it has been observed that certain sectors of the theory are more tractable.
One of them is the sector in which the states
 carry large angular momentum $\it{J}$ or
 large R-charge in the CFT point of view \cite{Berenstein:2002jq}.
  In this region, semi-classical string energies yield information on
the quantum spectrum of the string \cite{Gubser:2002tv}.
 The operator/string correspondence
then allows to associate dual long trace operators in
the gauge theory side to each string state.
 The energy eigenvalue
$E$ of these string states with respect to time
 in global coordinates is conjectured to be equal to the scaling dimension
 $\Delta$ of the dual gauge theory operators.
  This belief builds on a remarkable proposal
   \cite{Minahan:2002ve} relating the Hamiltonian of a Heisenberg's
spin chain system with that of the dilatation operator in N=4
supersymmetric Yang-Mills theory. Said another way,
 we have uncovered a very promising interplay
 between string theory, gauge theory ,and the spin-chain system.
 This could indeed pave the way for understanding new aspects of this field of research.

One of the most interesting properties of these  spin chain systems
is the so called magnon excitation. This was initiated in the work
of Hofman and Maldacena \cite{Hofman:2006xt} who showed that these
magnon states correspond to the  specific configuration of
semi-classical string states on $R\times S^2$ \cite{Gubser:2002tv}.
In particular, the giant magnon solution  corresponds to operators
where one of the $SO(6)$ charge, $J$, is taken to infinity, keeping
the $E-J$ fixed\footnote{The Hofman-Maldacena limit: $J\rightarrow
\infty, \lambda = {\rm fixed}, p= {\rm fixed}, E-J = {\rm fixed}$}.
These excitations satisfy a dispersion relation of the type (in the
large ' t Hooft limit $(\lambda)$)
\begin{equation}\label{EminJ}
 E-J = \frac{\sqrt{\lambda}}{\pi}
\left|{\rm sin}\frac{p}{2}\right| \ ,
\end{equation}
where $p$ is the magnon momentum. Hence after, a lot of work has been
devoted to study and generalize to various other magnon states with
two and three non vanishing momentum and so on \cite{Dorey:2006dq,Minahan:2006bd,Chen:2006ge,
Chu:2006ae,Spradlin:2006wk,Bobev:2006fg,Ryang:2006yq,Banerjee:2014rza,Gwak:2009nq,
Ahn:2008sk,Bozhilov:2007wn,Ahn:2008sk,Bozhilov:2007wn,Bobev:2007bm,Arutyunov:2006gs}.

As was stressed in \cite{Hoare:2014pna}, it is crucial to find the
gauge theory dual to string theory in the deformed geometry. One of
the modest steps would be to find the magnon-like dispersion
relation for strings moving in a given geometry and this is exactly the
goal of this paper. More explicitly, we analyze the motion of the
Polyakov string in the deformed geometry $AdS_3\times S^3$ whose
dynamics have a similar form as the dynamics of magnon-like string
on $R\times S^2$ \cite{Hofman:2006xt}. We analyze the equations of
motion that follow from the given action and the corresponding Virasoro
constraints. We consider infinite $\it{J}$ magnon solution when both $E$ and
$J$ diverge. Nevertheless, we construct a finite difference out of these two divergent quantities where this result has the form of the magnon-like dispersion relation. The latter has a complicated form which, however, reduces to the standard relation (\ref{EminJ}) in the limit
$\kappa\rightarrow 0$. This fact certainly justifies our
lacking of the corresponding dual gauge theory interpretation.

This paper is organized as follows. In section (\ref{second}),
we introduce the 3d truncated metric along with the Polyakov string action in this $\it{AdS_3\times S^3}$ background. We derive the corresponding equations of
motion and solve them with the specific magnon-like ansatz. In
section (\ref{third}),  we consider the limit of infinite angular momentum. A detailed manipulation is then presented leading to the giant magnon dispersion relation. Finally, in the conclusion
(\ref{fourth}), we outline our results and suggest possible extension of present work.

\section{Polyakov string in  deformed $AdS_3\times S^3$
background }\label{second}
 In this section, we study the Polyakov string in the
deformed $AdS_3\times S^3$ background that was  found recently in
\cite{Hoare:2014pna}. The line element has the following form
\begin{equation}\label{ds}
ds^2=ds^2_{A_3}+ds^2_{S^3} \ ,
\end{equation}
where
\begin{eqnarray}\label{dAS}
ds^2_{A_3}&=&-h(\rho)dt^2+f(\rho)d\rho^2+\rho^2d\psi^2 \ , \nonumber
\\
ds^2_{S^3}&=& \th(r)d\varphi^2+\tf(r)dr^2+r^2d\phi^2 \ , \nonumber \\
\end{eqnarray}
where
\begin{eqnarray}\label{defhf}
h&=&\frac{1+\rho^2}{1-\kappa^2\rho^2} \ , \quad
f=\frac{1}{(1+\rho^2)
(1-\kappa^2\rho^2)} \ , \nonumber \\
\th&=&\frac{1-r^2}{1+\kappa^2 r^2} \ , \quad \tf=\frac{1}{(1-r^2)
(1+\kappa^2 r^2)} \ . \nonumber \\
\end{eqnarray}
where the NS-NS  two form  vanishes.

We are interested in the geometry (\ref{dAS}) for finding out the
giant magnon solution that is analogue of the giant magnon solution
in $AdS_5 \times S^5$ found in \cite{Hofman:2006xt}. Our starting
point is the Polyakov form of the string action in  the deformed
background
\begin{eqnarray}\label{actPol}
S&=&-\frac{1}{2}\hT \int_{-\pi}^\pi d\sigma d\tau
\sqrt{-\gamma}\gamma^{\alpha\beta} g_{MN}\partial_\alpha
x^M\partial_\beta x^N  , \nonumber \\
\end{eqnarray}
where the effective string tension \cite{Hoare:2014pna}
 has the form
\begin{equation}
\hat{T}=\frac{\sqrt{\lambda}}{2\pi}\sqrt{1+\kappa^2} \ ,
\end{equation}
where the undeformed  string tension in $AdS$ background has the
famous form  $T_0=\frac{\sqrt{\lambda}}{2\pi}$. Further,
$\gamma^{\alpha\beta}$ is the world-sheet metric and
 the modes
$x^M, M=0,\dots,9$ parameterize the embedding of the string in the
general background.

The variation of the action
(\ref{actPol})
with respect to $x^M$ implies the
following equations of motion
\begin{eqnarray}
-\frac{1} {2}\sqrt{-\gamma}\gamma^{\alpha\beta}
\partial_K
g_{MN}\partial_\alpha x^M\partial_\beta x^N +\partial_\alpha[
\sqrt{-\gamma}\gamma^{\alpha\beta} g_{KM}\partial_\beta x^M]=0  \ .
\nonumber \\
\end{eqnarray} Further, the variation of the action with respect to
the metric implies the constraints
\begin{eqnarray}\label{gravcons}
T_{\alpha\beta}\equiv-\frac{2}{\sqrt{-\gamma}} \frac{\delta
S}{\delta \gamma^{\alpha \beta}}&=& \hT\left[ g_{MN}\partial_\alpha
x^M\partial_\beta x^N-\frac{1}{2} \gamma_{\alpha\beta}
\gamma^{\gamma\delta}
\partial_\gamma x^M\partial_\delta
x^N g_{MN}\right]=0 \ . \nonumber \\
\nonumber \\
\end{eqnarray}
Now, looking at the form of the background (\ref{dAS}), we observe that
the action (\ref{actPol})
 is invariant under the
following transformations of fields
\begin{eqnarray}
t'(\tau,\sigma)&=&t(\sigma,\tau)+\epsilon_t \ ,
\nonumber \\
\psi'(\tau,\sigma)&=& \psi(\tau,\sigma)+\epsilon_{\psi} \ ,
\nonumber \\
\phi'(\tau,\sigma)&=&\phi(\tau,\sigma)+\epsilon_{\phi} \ , \nonumber
\\
\varphi'(\tau,\sigma)&=&\varphi(\tau,\sigma)+\epsilon_{\varphi} \ ,
\nonumber \\
\end{eqnarray}
where $\epsilon_t,\epsilon_{\psi},\epsilon_{\phi}$ and
$\epsilon_{\varphi}$  are constants. Then, it is a simple task to
determine the corresponding conserved charges
\begin{eqnarray}\label{CGcon}
P_t&=&\hT \int_{-\pi}^\pi d\sigma \sqrt{-\gamma}\gamma^{\tau\alpha}
g_{tt}\partial_\alpha t \ , \nonumber \\
P_{\phi}&=& \hT \int_{-\pi}^\pi d\sigma
\sqrt{-\gamma}\gamma^{\tau\alpha} g_{\phi\phi}\partial_\alpha \phi \
,
\nonumber \\
P_{\psi}&=&\hT\int_{-\pi}^\pi d\sigma
\sqrt{-\gamma}\gamma^{\tau\alpha}g_{\psi\psi}\partial_\alpha \psi \
, \nonumber \\
P_{\varphi}&=&\hT\int_{-\pi}^\pi d\sigma
\sqrt{-\gamma}\gamma^{\tau\alpha}
g_{\varphi\varphi}\partial_\alpha\varphi \ . \nonumber \\
\end{eqnarray}
Note that $P_t$ is related to the energy as
$P_t=-E$.

Now we are going to
 find the solution of the equations of motion
given above which could be interpreted as a giant magnon. We closely
follow the very nice analysis presented in \cite{Arutyunov:2006gs}.

Let us now consider the following ansatz for obtaining the giant
magnon solution
\begin{eqnarray}\label{ans}
t=-\frac{E}{2\pi\hT} \tau \ , \quad
r=r(\sigma,\tau), \quad  \rho=\rho(\tau)\ , \nonumber \\
\varphi=\varphi(\sigma,\tau) \ , \quad \phi=const. \ , \psi=\Psi\tau
\ . \nonumber \\
\end{eqnarray}
 We will solve the equations of motions in the conformal
 gauge where $\gamma^{\tau\tau}=-1 \ ,
 \gamma^{\sigma\sigma}=1 \ , \gamma^{\tau\sigma}=0$.
Note that in the given gauge, the constraints
 (\ref{gravcons}) take  the form
\begin{eqnarray}\label{Tcon}
T_{\sigma\sigma}&=& \frac{\hT}{2} (g_{MN}\partial_\sigma
x^M\partial_\sigma x^N +g_{MN}\partial_\tau x^M\partial_\tau x^N) \
,
\nonumber \\
T_{\tau\tau}&=& \frac{\hT}{2} (g_{MN}\partial_\sigma
x^M\partial_\sigma x^N +g_{MN}\partial_\tau x^M\partial_\tau x^N)\ ,
\nonumber \\
T_{\tau\sigma}&=& \hT g_{MN}\partial_\sigma x^M\partial_\tau x^N
 \ .
\nonumber \\
\end{eqnarray}
Let us now consider the equation of motion for $X^0\equiv t$
\begin{eqnarray}
\partial_\alpha[\sqrt{-\eta}\eta^{\alpha\beta}
g_{tt}\partial_\beta t]=0
\nonumber \\
\end{eqnarray}
that for the ansatz  (\ref{ans}) implies that $\rho$ should be
constant $\rho_c$.
On the other hand, the equation of motion with respect to $\rho$ has
the form
\begin{eqnarray}
\frac{\rho(1+\kappa^2)}{(1-\kappa^2\rho^2)^2}-\rho \Psi^2=0 \
\nonumber \\
\end{eqnarray}
that has solution $\rho_c=0$ and hence we find $g_{tt}(\rho_c=0)=1$.
 As a check, note that when we insert (\ref{ans}) into the first
 expression in (\ref{CGcon}) and use $g_{tt}=1$ we find
\begin{equation}
P_t=-\hat{T}\int_{-\pi}^\pi d\sigma  \frac{E}{2\pi\hT}=-E \ .
\end{equation}
Now, we return to the equation of motion for $\phi$
\begin{equation}
\partial_\alpha[g_{\phi\phi}\eta^{\alpha\beta}
\partial_\beta\phi]=0
\end{equation}
and we see that it is solved form $\phi=\mathrm{const}$. Finally,
the equation of motion for $\psi$ is obeyed with (\ref{ans}) since
$\rho_c=0$. Then we also have $P_\psi=0$.

Now, we proceed to the analysis of the dynamics of $r$ and $\varphi$.
It is convenient to  solve the constraints (\ref{Tcon}) that can be
interpreted as the first integrals of the theory instead of
solving the equations of motion for $r$ and $\varphi$
respectively. Inserting
 the ansatz (\ref{ans}) to   the constraints
 (\ref{Tcon}), we obtain two
equations
\begin{eqnarray}\label{Tconas}
 T_{\sigma\sigma}&=&T_{\tau\tau}=\hT\left[
-\frac{1}{(2\pi \hT)^2}E^2 +\tf(r) (\partial_\tau
r)^2+\tf(r)(\partial_\sigma r)^2+\th(r)(\partial_\tau \varphi)^2+
\th(r)(\partial_\sigma \varphi)^2\right]=0 \ ,  \nonumber \\
T_{\tau\sigma}&=&\hT \left[\tf(r)\partial_\tau r
\partial_\sigma r+\th(r)\partial_\tau \varphi\partial_\sigma
\varphi \right]=0 \ . \nonumber \\
\end{eqnarray}
 Following \cite{Hofman:2006xt,Arutyunov:2006gs},
 we   search for a
solution with the boundary conditions
\begin{equation}
r(\pi,\tau)-r(-\pi,\tau)=0 \ , \quad \triangle \varphi=
\varphi(\pi,\tau)-\varphi(-\pi,\tau)=p \ ,
\end{equation}
where $p$ is the momentum of the `single magnon' excitation. Since
the field $\varphi$ does not satisfy periodic boundary conditions,
this solution corresponds to the open string.

As the next step we introduce the light-cone
coordinate $\varphi$ through the formula
\begin{equation}
\varphi=\tilde{\varphi}+\omega \tau
\end{equation}
and presume that  $\tilde{\varphi}$ and $r$ depend on $\tau$ and
$\sigma$ in the following way
\begin{equation}
r=r(\sigma-v\omega \tau) \ ,
\quad \tilde{\varphi}=\tilde{\varphi}(\sigma-v\omega \tau) \ .
\end{equation}
Inserting these forms into  (\ref{Tconas}),  we obtain
\begin{eqnarray}\label{Virins}
& &-v\omega \tf (r)r'^2+\th(r)
(\omega-v\omega\tilde{\varphi}')\tilde{\varphi}'=0 \ ,
\nonumber \\
& & -\mC
+\tf(r)r'^2(1+v^2\omega^2)+\th(r)(\omega-v\omega\tilde{\varphi}')^2+
\th(r)\varphi'^2
=0 \ ,  \nonumber \\
\end{eqnarray}
where
\begin{equation}
\mC=\frac{1}{(2\pi \hT)^2} E^2 \ ,
\end{equation}
and where now $r',\tilde{\varphi'}$ mean derivatives with respect to
$\xi\equiv\sigma-v\omega\tau$. If we now combine these equations we
obtain
\begin{eqnarray}\label{sol}
\tilde{\varphi'}&=&\frac{v}{\th(r)(1-v^2\omega^2)}\left(\mC
-\th(r)\omega^2\right) \ ,
\nonumber \\
r'^2&=&\frac{\omega^2}{\tf(r)\th(r)(1-v^2\omega^2)^2}
\left(\frac{\mC}{\omega^2} -\th(r)\right)\left(\th(r)-v^2 \mC\right)
\ .
\nonumber \\
\end{eqnarray}
As a check, note that for $\kappa=0$ we have $\th(r)= 1-r^2 \ ,
\tf(r)=\frac{1}{1-r^2}$. Then, we introduce $\theta$ as
$r=\cos\theta$ and we obtain
\begin{equation}
\tilde{\varphi'}=\frac{v}{(1-v^2\omega^2)\sin^2\theta}
(\mC-\omega^2\sin^2\theta)
\end{equation}
and also
\begin{equation}
\theta'^2=\frac{1}{\sin^2\theta}r'^2=\frac{\omega^2}{
(1-v^2\omega^2)^2\sin^2\theta}\left(\frac{\mC}{\omega^2}-\sin^2\theta\right)
\left(\sin^2\theta-v^2\mC\right) \
\end{equation}
that has the same form as the equation that determines giant magnon
profile that was derived in
 \cite{Arutyunov:2006gs}.
We also  see, in agreement with \cite{Arutyunov:2006gs},  that for
this solution the derivative $r'$ is finite everywhere and vanishes at both points $r_{min}$ and $r_{max}$ defined as\footnote{
 These conditions
follow from the requirement that $\frac{\mC}{\omega^2}>\th(r) $ and
$\th(r)- v^2\mC>0$ in order to have $r'^2>0$. The opposite condition
$\frac{\mC}{\omega^2}<\th(r) $ and $\th(r)- v^2\mC<0$ cannot be
obeyed when we presume $v<\frac{1}{\omega}$.}

\begin{eqnarray}\label{defrminrmax}
r_{min}=\sqrt{\frac{1-\frac{\mC}{\omega^2}}{1+\frac{\kappa^2}{\omega^2}\mC}}
\ , \quad  r_{max}=\sqrt{\frac{1-v^2\mC}{1+v^2\kappa^2\mC}} \ ,
\nonumber
\\
\end{eqnarray}
where we presume $v<\frac{1}{\omega}$.

As the next step, we insert (\ref{sol}) into definition of the
conserved charge $P_{\varphi}$ and we obtain
\begin{eqnarray}
P_{\varphi}=J=
2\hT \int_{r_{min}}^{r_{max}}
 \frac{\omega}{|r'|}
 \th(r)
 (v\tilde{\varphi}'-1)
=-2\hT\frac{\omega}{|\omega|}\sqrt{\frac{ (1+v^2\mC\kappa^2)} {
(\frac{\mC}{\omega^2}\kappa^2+1)}}
\times \nonumber\\
\times \int_{r_{min}}^{r_{max}} \frac{dr}
{(1+\kappa^2r^2)\sqrt{(r^2-r^2_{min})(r^2_{max}-r^2)}}
(r_{max}^2-r^2) \ ,  \nonumber \\
\end{eqnarray}
Note that we also  have
\begin{equation}
\begin{split}
2\pi& = \int_{-\pi}^\pi d\sigma=
 2\int_{r_{min}}^{r_{max}}\frac{dr}{|r'|}=
 2\frac{(1-v^2\omega^2)}{\omega\sqrt{(1+\frac{\mC}{\omega^2}\kappa^2)(1+v^2\kappa^2\mC)}}\times\\
& \times
\int_{r_{min}}^{r_{max}}\frac{dr}{\sqrt{(r^2_{max}-r^2)(r^2-r^2_{min})}}
\ .
\end{split}
\end{equation}
Finally we also find
\begin{eqnarray}\label{ptriangle}
\begin{split}
p & = \triangle \varphi=2 \int_{r_{min}}^{r_{max}}\frac{dr}{|r'|}
\tilde{\varphi}'=  \\
& =
2v\frac{\omega^2}{|\omega|}\sqrt{\frac{1+\frac{\mC}{\omega^2}\kappa^2}{
1+\kappa^2\mC v^2}}
\int_{r_{min}}^{r_{max}}\frac{dr}{1-r^2}\frac{r^2-r_{min}^2}
{\sqrt{(r^2-r_{min}^2)(r_{max}^2-r^2)}} \ .
\end{split}
\end{eqnarray}
In principle we could express these quantities  in terms of
elliptical integrals. However, in order to derive more transparent
results, we restrict ourselves to the case of the infinite giant
magnon, where $J\rightarrow \infty$.
\section{Infinite $J$ Giant Magnon}\label{third}
In this section, we discuss the infinite $\it{J}$ giant magnon.
 This solution  corresponds to the situation when  $r_{min}\rightarrow 0$. From
(\ref{defrminrmax}), we find that it occurs when
\begin{equation}
\omega^2=\mC \ , r_{max}=\sqrt{\frac{1-v^2\omega^2}{
1+v^2\omega^2\kappa^2}} \ .
\end{equation}

Let us calculate $P_{\varphi}$ for $r_{min}=0$
\begin{eqnarray}
P_{\varphi}&=&2\hT\frac{\omega r_{max}}{|\omega|}
\sqrt{\frac{1+v^2\mC\kappa^2}{ 1+\frac{\mC}{\omega^2}\kappa^2}}
\left[\frac{\sqrt{1+\kappa^2r_{max}^2}}{\kappa r_{max}}
\ln\sqrt{\frac{1+\frac{\kappa
r_{max}}{\sqrt{1+\kappa^2r_{max}^2}}}{1-
\frac{\kappa r_{max}}{\sqrt{1+\kappa^2r_{max}^2}}}}+\right.\nonumber \\
& &\left. +\ln
\sqrt{\frac{1+\sqrt{1-(r/r_{max})^2}}{1-\sqrt{1-(r/r_{max})^2}}}|_0^{r_{max}}
\right]
\nonumber \\
\end{eqnarray}
and we see that the second expression diverges as anticipated.
However, the same divergence occurs when we calculate the integral
\begin{eqnarray}\label{2piinf}
\begin{split}
2\pi & = \int_{-\pi}^\pi d\sigma= 2\int_{r_{min}}^{r_{max}}\frac{dr}{|r'|} \\
& =
-2\frac{(1-v^2\omega^2)}{|\omega|\sqrt{(1+\frac{\mC}{\omega^2}\kappa^2)(1+v^2\kappa^2\mC)}}
\frac{1}{r_{max}}\ln \sqrt{\frac{1+\sqrt{1-(r/r_{max})^2}}
{1-\sqrt{1-(r/r_{max})^2}}}|_0^{r_{max}} \
\end{split}
\end{eqnarray}
so that it is natural to regularize the divergence in $P_{\varphi}$
using (\ref{2piinf}) in order to find dispersion relation that would
be analogue of (\ref{EminJ}).
Explicitly, let us consider the following combination $2\pi
K-P_{\varphi}$ where we choose the constant $K$ to make this
difference finite. It is easy to see that this divergence is
canceled out when $K$ is equal to
\begin{equation}
K=-\hT \omega \ .
\end{equation}
To proceed further, we  choose $\omega=-\sqrt{\mC}=-\frac{E}{2\pi\hT}$ in order to have positive charge
$J$. Then, we finally find
\begin{equation}
K=\frac{E}{2\pi}
\end{equation}
and hence we have the following dispersion relation
\begin{eqnarray}\label{helpdis}
 E
 -J
= \frac{2\hT}{\kappa}\sqrt{\frac{1+v^2\mC\kappa^2}{
1+\frac{\mC}{\omega^2}\kappa^2}} \sqrt{1+\kappa^2r_{max}^2}
\ln\sqrt{\frac{1+\frac{\kappa
r_{max}}{\sqrt{1+\kappa^2r_{max}^2}}}{1- \frac{\kappa
r_{max}}{\sqrt{1+\kappa^2r_{max}^2}}}} \ .
\end{eqnarray}
Finally, we  calculate the momentum defined in (\ref{ptriangle}) for
$r_{min}=0$
\begin{eqnarray}\label{ptriangleinf}
p=2v\frac{\omega^2}{|\omega|}
\sqrt{\frac{(1+\frac{\mC}{\omega^2}\kappa^2)} {1+\kappa^2\mC v^2}}
\int_0^{r_{max}}dr \frac{r}{(1-r^2)\sqrt{r^2_{max}- r^2}}
=2\sin^{-1}r_{max}
\end{eqnarray}
so that
\begin{equation}
r_{max}=\sin \frac{p}{2} \ .
\end{equation}
Note that using this relation we can express $v^2\mC$ as function of
$p$
\begin{equation}
v^2\mC=\frac{1-\sin^2\frac{p}{2}}{1+\kappa^2\sin^2\frac{p}{2}} \ .
\end{equation}
 Plugging this back into (\ref{helpdis}) we find the final form of the
 giant magnon dispersion relation
\begin{eqnarray}\label{disprelation}
E
 -J=
\frac{2\hT}{\kappa} \ln\sqrt{\frac{1+\frac{\kappa
|\sin\frac{p}{2}|}{\sqrt{1+\kappa^2\sin^2\frac{p}{2}}}}{1-
\frac{\kappa
|\sin\frac{p}{2}|}{\sqrt{1+\kappa^2\sin^2\frac{p}{2}}}}}=
\frac{2\hT}{\kappa}\tanh^{-1}\left(\frac{\kappa |\sin\frac{p}{2}|}{
\sqrt{1+\kappa^2 \sin^2\frac{p}{2}}}\right) \ .
\nonumber \\
\end{eqnarray}
Obviously, we see that it has a much more complicated form than in
the case of $\kappa=0$. At present it is not clear  how this
dispersion relation could be realized in the dual gauge theory
description at least in some limit. On the other hand, it is
instructive to study the limit  $ \kappa \rightarrow 0$ in
(\ref{disprelation}). In fact, using
\begin{equation}
\lim_{\kappa\rightarrow 0}\frac{1}{\kappa}
\ln\sqrt{\frac{1+\frac{\kappa |
\sin\frac{p}{2}|}{\sqrt{1+\kappa^2\sin^2\frac{p}{2}}}}{1-
\frac{\kappa
|\sin\frac{p}{2}|}{\sqrt{1+\kappa^2\sin^2\frac{p}{2}}}}}
=\left|\sin\frac{p}{2}\right|
\end{equation}
 we find that (\ref{disprelation}) takes the form
\begin{equation}
E-J=2\hT(\kappa=0)\sin\frac{p}{2}=\frac{\sqrt{\lambda}}{\pi}\left|\sin\frac{p}{2}\right|
\
\end{equation}
in exact agreement with \cite{Hofman:2006xt}. Further, using the
fact that $\kappa$ is related to the deformations parameters
introduced in \cite{Arutyunov:2013ega}
\begin{equation}
\kappa=\frac{2\eta}{1-\eta^2} \ , \quad q=e^{-\frac{\nu}{T_0}} \ ,
\quad \nu=\frac{2\eta}{1+\eta^2}=\frac{\kappa}{\sqrt{1+\kappa^2}} \
\end{equation}
we find that (\ref{disprelation}) can be written as
\begin{equation}
E-J=\frac{2T_0}{\nu}\sinh^{-1}\left(\frac{\nu}{\sqrt{1-\nu^2}}\left|\sin\frac{p}{2}\right|\right)
\
\end{equation}
and we see that the right side of the equation above precisely agree
with the  energy of the magnon excitation calculated in
\cite{Arutynov:2014ota}. This fact can be considered as further
support of our analysis.
\section{Conclusion}\label{fourth}
In this paper, we have studied the giant magnon solution in deformed
$AdS_3\times S^3$  background that was proposed recently in
\cite{Hoare:2014pna}.
 In the infinite  $J$ limit we
were able to compute the one magnon dispersion relation that has
slightly more complicated form  than the ordinary dispersion
relation in case of giant magnon in $AdS_5\times S^5$. However, we
have further argued that this dispersion relation coincides with the
latter in the limit $\kappa\rightarrow 0$.

The present analysis can be extended in various directions.
 In one event, one
 can ask what role these results play on the gauge theory
  side and what interpretation they provide.
   The deformed string metric (\ref{ds})
   has a curvature singularity at $\rho = \frac{1}{\kappa}$.
   For larger  values, the radial coordinate $\rho$ becomes
   time-like that suggests that string is confined in the region
   $0\leq \rho\leq \frac{1}{\kappa}$.
Then it is not completely clear  how to proceed with our results on
the dual gauge side. In other words,  it would  be extremely
interesting to understand in detail the holographic duality between
string theory in the deformed geometry  and the dual  gauge theory.
 Clearly, this is an interesting topic for further
work. There are more other lines of investigation worth pursuing.
For example, it would be very interesting and challenging to include
the 2-form $B$ field into our calculations when we consider the
giant magnon solution on the deformed $AdS_5\times S^5$ background.
It would be also interesting to analyze corresponding spike
solutions in given background. These problems are currently under
investigations.

\vskip .5in \noindent {\bf Acknowledgement:}
\\
We would like to thank Stijn van Tongeren for his important remark
about our result and the analysis performed in
\cite{Arutynov:2014ota}.
  This work   was
supported by the Grant agency of the Czech republic under the grant
P201/12/G028.

\end{document}